# Origin of the Anomalous Electronic Shot Noise in Atomic-Scale Junctions

Anqi Mu,[a,†] Ofir Shein Lumbroso[a,‡] Oren Tal,[‡] and Dvira Segal[*,†]

[†]*Department of Chemistry and Centre for Quantum Information and Quantum Control, University of Toronto, 80 Saint George St., Toronto, Ontario, Canada M5S 3H6*

[‡]*Department of Chemical and Biological Physics, Weizmann Institute of Science, Rehovot, Israel*

E-mail: dvira.segal@utoronto.ca

[a]A.M. and O.S.L. contributed equally to this work

July 16, 2019

## Abstract

Fluctuations pose fundamental limitations in making sensitive measurements, yet at the same time, noise unravels properties that are inaccessible at the level of the averaged signal. In electronic devices, shot noise arises from the discrete nature of charge carriers and it increases linearly with the applied voltage according to the celebrated Schottky formula. Nonetheless, measurements of shot noise in atomic-scale junctions at high voltage reveal significant nonlinear (anomalous) behavior, which varies from sample to sample, and has no specific trend. Here, we provide a viable, unifying explanation for these diverse observations based on the theory of quantum coherent transport. Our formula for the anomalous shot noise relies on—and allows us to resolve—two key characteristics of a conducting junction: The structure of the transmission function at the vicinity of the Fermi energy and the asymmetry of the bias voltage drop at the contacts. We tested our theory on high voltage shot noise measurements on Au atomic scale junctions and demonstrated a quantitative agreement, recovering both the enhancement and suppression of shot noise as observed in different junctions. The good theory-experiment correspondence supports our modelling and emphasizes that the asymmetry of the bias drop on the contacts is a key factor in nanoscale electronic transport, which may substantially impact electronic signals even in incomplex structures.

# 1 Introduction

Noise in electronic signals is typically undesired, yet it can be a source of information on the conducting system by exposing effects concealed in the time-averaged electric current.[1–3] Shot noise measurements at the mesoscale and nanoscale[4,5] reveal the fractional charge of quasiparticles in many-body systems,[6] contributions of different conduction channels to the electronic transport,[7–13] the crossover from ballistic to diffusive transport,[14] the valence orbital structure at the contact,[15,16] activation of vibrations in molecular conducting junctions,[10,17–19] and the onset of spin-polarized transport.[20,21]

Focusing on the white noise (flat power spectrum) component, we recall on the different noise sources:[1,2] The thermal motion of charge carriers in electronic conductors is responsible for the Johnson-Nyquist noise,[22,23] which is proportional to the temperature and the linear response electrical conductance. When a voltage bias is applied across a conductor, voltage-induced shot noise is activated, and it dominates over the thermal noise at high bias and low temperatures. Temperature differences across a conductor activate an additional contribution, the recently identified 'delta-T' noise, which is quadratic in the temperature difference.[24]

Recent measurements of shot noise in atomic-scale junctions of Au revealed highly non-linear behavior of the noise as a function of bias voltage at high voltage and low temperature, which varies between samples.[25] These observations, as well as other measurements at el-evated temperatures[26,27] are anomalous in the sense that they do not follow the standard theoretical prediction [see Eqs. (14) and (15) below]. We recall that to derive the standard formulae, the transmission probability of electrons to cross an atomic-scale constriction is assumed to be constant (independent of energy and voltage), evaluated at the equilibrium Fermi energy.[2]

Different mechanisms were suggested to explain the observation of anomalous shot noise: local heating of electrons,[26] quantum interference due to scatterings with impurities located at the electrodes at the vicinity of the point contact,[25] electron-electron and electron-phonon inelastic effects.[27] Correspondingly, recent theoretical works focused on the behavior of shot noise while taking into account electron,[28–30] and spin[31] correlations, as well as inelastic electron-phonon effects.[32–36] Approximately, the impact of such processes can be captured within the elastic transport theory by using a voltage, and/or temperature dependent transmission function.[25,26] Nevertheless, a consistent-rigorous theoretical explanation to the variety of experimental observations of anomalous shot noise is still missing, even for the very well-studied metallic Au break junction setup.[25–27]

The objective of our work is to derive a closed-form formula for the electronic shot noise in atomic-scale conductors, which reaches the far from equilibrium (high bias voltage) regime. Recent studies have put forward complex mechanisms for explaining observations of anomalous shot noise, e.g. relying on many body effects. In contrast, here we carefully examine the analytically tractable problem of elastic conduction in quantum coherent junctions and bring to light a nonlinear behavior that builds up far from equilibrium. Our formula for the high-voltage shot noise takes into account two key factors: (i) The transmission function varies with energy. This dependence could stem from different underlying microscopic effects such as the occurrence of electronic resonances or scattering of carriers at and near the contacts due to defects. (ii) The bias voltage may drop unevenly on the contacts.

While quantum chemistry computations can be readily performed to take into account the rich electronic structure of a specific atomic or molecular junction, as was recently done in Refs.,[37,38] our goal here has been to portray a more universal picture of quantum coherent shot noise. Therefore, our formula for the anomalous shot noise does not assume a specific mechanism underlying the energy dependent transmission function—yet it allows us to deduce on fundamental parameters characterizing the conducting junction.

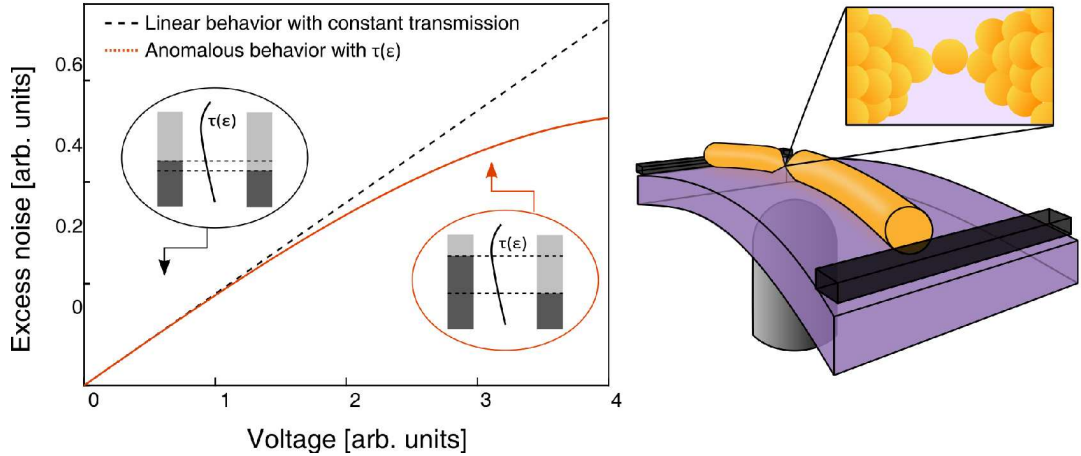

Figure 1: An illustrative example of nonlinear shot noise examined in this work, along with a sketch of the mechanically-controllable break junction setup that was used to test our theoretical derivations, with an atomic-scale Au junction zoomed in. When the applied voltage is small, the assumption of a constant transmission function is justified (dashed). Under high bias, the energy dependency of the transmission function should be considered (full). Since the atomic configuration at the junction may be asymmetric, as we illustrate here, the applied voltage may be partitioned unevenly across the atomic-scale junction.

We report on anomalous shot noise measurements in Au atomic-scale junctions as sketched in Fig. 1, and test our theory on different realizations of these junctions that show distinct nonlinear trends. By analyzing hand in hand the differential conductance and the *linear* (low bias) region of the shot noise, we are able to quantitatively reproduce the experimental results for the *anomalous* (high bias) shot noise using our formulation without fitting parameters, and in variety of junctions construing either a suppression or an enhancement of shot noise at high voltage. The theory-experiment agreement well supports our physical picture and modelling. Moreover, our analysis allows us to estimate the lower bound for the asymmetry of voltage drop on the junction and the energy variation of the transmission function at the vicinity of the Fermi energy.

# 2 Results and Discussion

## 2.1 Theory of the anomalous shot noise

From the theory of coherent elastic transport we derive here a closed-form formula for the nonlinear shot noise, given in terms of the energy dependent transmission function and bias drop asymmetry. We consider a two-terminal junction under applied bias voltage $\Delta\mu = eV$, at a fixed temperature $T$. Ignoring decoherence and inelastic processes within the atomic-scale constriction, the average current is given by the Landauer formula, which is written in terms of the transmission function $\tau(\epsilon)$ with $\epsilon$ the energy of incoming charge carriers. The transmission function describes the probability for an incoming electron at a given energy to be transmitted across the junction. In resonant transport, the transmission function is peaked at energies of charge conducting atomic or molecular orbitals (resonances) and its width reflects the hybridization strength of that state with the electrodes' frontier states.[39] Furthermore, the transmission function may depend on energy due to quantum interference effects with defects in the contacts at the vicinity of the atomically-defined junction. The standard assumption of a constant transmission function is justified at low bias voltage and under low temperatures. Then, the transmission function can be approximated by its (fixed) value at the Fermi energy, see Fig. 1.

Our derivation of the anomalous shot noise combines two elements (Section S1, Supporting Information). First, since we are interested in high-voltage effects, we take into account the variation of the transmission function with energy around the equilibrium Fermi energy $\mu$, see Fig. 1. As a simple approximation, we expand the transmission as a Taylor series to the lowest order, $\tau(\epsilon) \approx \tau_0 + \tau'(\mu)(\epsilon - \mu)$, with $\tau_0$ the transmission probability at the Fermi energy. In this expansion, the transmission function is assumed to be linear in energy around $\mu$. Second, the bias voltage is (possibly) divided asymmetrically at the contacts. This asymmetry is quantified by the parameter $0 \leq \alpha \leq 1$ such that $\mu_L = \mu + \alpha\Delta\mu$, $\mu_R = \mu - (1-\alpha)\Delta\mu$. If $\alpha = \frac{1}{2}$, the potential is symmetrically divided at the electrodes, $\mu_{L,R} = \mu \pm \Delta\mu/2$. When $\alpha = 1$ (0) the bias entirely falls on the left (right) electrode. The origin of this asymmetry could be structural differences in the contact region at the left and right sides, e.g., the atomic configuration at the contact region could be somewhat different. The parameter $\alpha$ is rooted in many-body effects: The potential distribution across the junction is generated by the nonequilibrium situation, with charges reorganized due to bias voltage. Therefore, at high bias the electric potential in the junction should be determined in a self consistent manner.[40,41] Here, following other studies, for example,[32] we use a non-interacting electron formalism, but account for such a many-body effect by introducing the $\alpha$ parameter that captures the bias drop on the junction, without performing a detailed self-consistent electronic structure calculation.

Using these two elements in our modelling, the averaged charge current is obtained from the Landauer formula as (see Methods Section),[42]

$$\langle I \rangle = \frac{2e}{h}\tau_0 \Delta\mu + \frac{2e}{h}\tau'(\mu)\left(\alpha - \frac{1}{2}\right)(\Delta\mu)^2. \quad (1)$$

For simplicity, we consider here a single transmission channel. Notably, the quadratic $(\Delta\mu)^2$ term appears only if the bias drops asymmetrically on the contacts.

We proceed and evaluate the formally-exact integral expression for the shot noise in coherent quantum conductors [see Eq. (13) in Methods section][2] using the linear expansion for the transmission function. Our result, a closed-form expression for the voltage-activated shot noise at nonzero temperature $T$ and at arbitrary voltage, can be decomposed into three terms (Sections S2-S3, Supporting Information),

$$S^T_{\Delta\mu} = S_n + S_a + S_a^{\alpha \neq 1/2}. \quad (2)$$

The "normal" shot noise is given by the standard formula,[2,45]

$$S_n = 4k_B T \tau_0 G_0$$
$$+ 4k_B T G_0 \left[ \frac{\Delta \mu}{2k_B T} \coth\left(\frac{\Delta \mu}{2k_B T}\right) - 1 \right] \tau_0 (1 - \tau_0). \tag{3}$$

The other two contributions in Eq. (2) are termed "anomalous". When the voltage bias is perfectly symmetric, the nonlinear shot noise is solely given by $S_a$ while $S_a^{\alpha \neq 1/2}$ precisely cancels. In contrast, in the case of the asymmetric voltage drop, both anomalous terms contribute, but $S_a^{\alpha \neq 1/2}$ dominates the noise, as we discuss below. Explicitly,

$$S_a = -4k_B T G_0 \left[ \frac{\Delta\mu}{2k_B T} \coth\left(\frac{\Delta\mu}{2k_B T}\right) - 1 \right] \left[ [\tau'(\mu)]^2 \frac{\pi^2 k_B^2 T^2}{3} + [\tau'(\mu)]^2 \frac{(\Delta\mu)^2}{12} \right]$$
$$+ \frac{2}{3} k_B T G_0 [\tau'(\mu)]^2 (\Delta\mu)^2,$$
$$S_a^{\alpha \neq 1/2} = 8 G_0 k_B T \tau_0 [\tau'(\mu)] \Delta\mu \left(\alpha - \frac{1}{2}\right) + 4 G_0 k_B T [\tau'(\mu)]^2 \left(\alpha - \frac{1}{2}\right)^2 (\Delta\mu)^2$$
$$+ 2 G_0 \coth\left(\frac{\Delta\mu}{2k_B T}\right) \left[ (1 - 2\tau_0)(\Delta\mu)^2 \left(\alpha - \frac{1}{2}\right) [\tau'(\mu)] - \left(\alpha - \frac{1}{2}\right)^2 (\Delta\mu)^3 [\tau'(\mu)]^2 \right]. \tag{4}$$

Equation (2) with the anomalous contributions (4) is a central result of our work. It is exact—to the order of $\tau'(\mu)$ considered. Obviously, the anomalous terms are cumbersome, and to gain insight we simplify these expressions in two scenarios. First, focusing on the strictly symmetric bias drop case, $\alpha = \frac{1}{2}$, we note that $S_a$ scales as $(\tau')^2$. The leading anomalous correction at low temperature is cubic in voltage, and the total noise is

$$S_{\Delta\mu}^{T \to 0} \approx S_n - G_0 [\tau'(\mu)]^2 \frac{|(\Delta\mu)^3|}{6}. \tag{5}$$

In contrast, $S_a^{\alpha \neq 1/2}$ has terms linear in $\tau'(\mu)$, which is our perturbative parameter. Therefore, assuming $[\tau'(\mu)]\Delta\mu$ to be small, once voltage-drop imbalance is permitted with $\alpha \neq \frac{1}{2}$, the nonlinear term $S_a^{\alpha \neq 1/2}$ outweighs $S_a$, and we obtain a distinct expression for the total shot noise,

$$S_{\Delta\mu}^{T\to 0} \approx S_n + 2G_0(1-2\tau_0)\tau'(\mu)\left(\alpha - \frac{1}{2}\right)(\Delta\mu)^2 - 2G_0[\tau'(\mu)]^2\left(\alpha - \frac{1}{2}\right)^2 |(\Delta\mu)|^3. \quad (6)$$

Consistent with our working assumption of slowly-varying transmission function, it is reasonable to neglect the last term in Eq. (6). It is significant to note that the combination $\tau'(\mu)(\alpha - 1/2)$ appears in both the expression for the current (1) and the noise, and it can be extracted from experimental data as we demonstrate below. Equation (5) predicts noise suppression at high bias. Therefore, if we observe an enhancement of shot noise beyond $S_n$, we can immediately conclude that the bias drops asymmetrically on the junction.

We retain the full temperature dependence of the standard shot noise contribution $S_n$ in Eqs. (5) and (6) so as to properly interpolate the noise formula to the $\Delta\mu \to 0$ limit. When several channels are involved in the transport process, Eqs. (1)-(6) are trivially generalized: since we work under the assumption of elastic transport, each channel independently-additively contributes.

Altogether, our analytical results illustrate that shot noise becomes nonlinear at high voltage once we take into account the energy dependence of the transmission function. Our results, Equations (1) - (6), were derived under the first-order Taylor expansion of the transmission function. Complementary formulae that are based on a quadratic expansion of the transmission function are also derived (section S4, Supporting Information).

Next, we use simulations and new experimental data to assess the validity of our analytical results. For simplicity, with simulations we probe the symmetric setup, $\alpha = 1/2$, Eq. (5). The more rampant situation with bias drop asymmetry, Eq. (6), is applied onto experimental data. Additional experimental results are presented in section S4, Supporting Information.
.

## 2.2 Simulations

As a specific example for an energy dependent transmission function, we consider a central model in molecular electronic transport, that is a resonant level with a single orbital at energy $\epsilon_d$ and a broadening $\Gamma_{L,R}$, arising due to its coupling to the left and right metals,

$$\tau(\epsilon) = \frac{\Gamma_L \Gamma_R}{(\epsilon - \epsilon_d)^2 + (\Gamma_L + \Gamma_R)^2/4}. \tag{7}$$

This Lorentzian function can be approximated by the linear expansion $\tau(\epsilon) \approx \tau_0 + \tau'(\mu)(\epsilon - \mu)$ with

$$\begin{aligned} \tau_0 &= \frac{\Gamma_L \Gamma_R}{(\epsilon_d - \mu)^2 + (\Gamma_L + \Gamma_R)^2/4}, \\ \tau'(\mu) &= \frac{2(\epsilon_d - \mu)\Gamma_L \Gamma_R}{[(\mu - \epsilon_d)^2 + (\Gamma_L + \Gamma_R)^2/4]^2}, \end{aligned} \tag{8}$$

as long as the broadening of the resonance is large and the resonance is placed aside of the Fermi energy, $\epsilon_d - \mu \neq 0$. In fact, the current noise can be evaluated analytically with the Lorentzian form,[32] but here our goal has been to test the simple formula (5), which should hold for broad resonances. In simulations we assume perfect spatial symmetry, $\Gamma = \Gamma_{L,R}$ and a symmetric bias drop, $\alpha = 1/2$. Numerical results (based on the integral expression, Eq. (13)) are compared to the constant transmission expression, $S_n$, and to the anomalous shot noise closed-form formula, (5).

The excess current noise includes contributions beyond the Johnson-Nyquist thermal noise, $S_{\Delta\mu}^T - S_{\Delta\mu=0}^T$. In Fig. 2 we present the excess noise at low temperature using a broad transmission function centered at $\epsilon_d = 0.4$ eV; the Fermi energy is set at zero. Our approach is based on the assumption that $\tau(\epsilon)$ slowly varies with energy in the relevant bias window. The current-voltage characteristics displayed in panel (b) is quite linear in the full range, which is expected under symmetric splitting of the applied voltage since $\langle I \rangle = \tau_0 G_0 \Delta\mu$. The current noise is linear in $V$ at low voltage: The thermal energy is $k_B T \sim 1$ meV for $T = 10$ K, and at low voltage (here up to 100 mV) the noise follows the simple formula $S_{\Delta\mu} = 2|\Delta\mu|GF$, with the Fano factor $F = \frac{\tau_0(1-\tau_0)}{\tau_0}$. However, at high bias the current noise, as computed numerically [see Eq. (13) in Methods], clearly deviates from the linear trend predicted by Eq. (3), while Eq. (5) provides an excellent match up to $\Delta\mu \sim 0.6$ eV. This agreement is substantial given that at this point the bias window covers a large portion of the transmission function, see Fig. 2. Simulations were performed at 10 K, but similar results were observed for a range of temperatures, $T$=0.1-300 K.

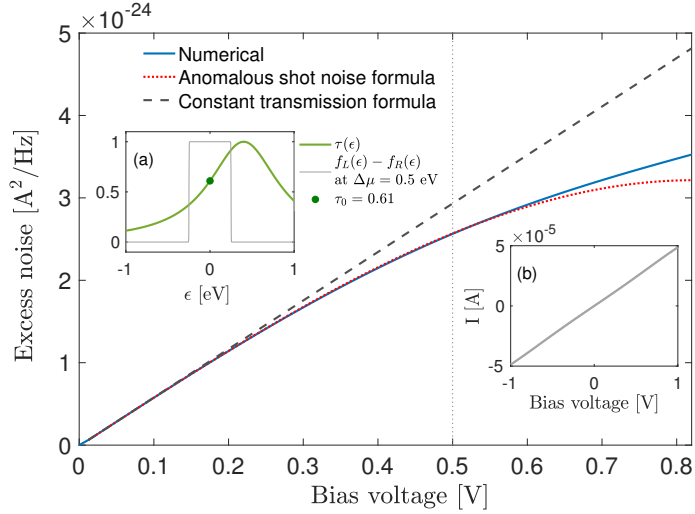

Figure 2: Simulations of excess current noise in an atomic-scale junction with a broad resonance. (a) Lorentzian transmission function located at $\epsilon_d = 0.4$ eV with $\Gamma = 0.5$ eV. We further present the Fermi function window at $\Delta\mu = 0.5$ eV and mark the transmission value at the Fermi energy, $\tau_0 = 0.61$. (b) Current-voltage characteristics using the transmission function from panel (a). In the main panel we display the excess current noise based on the numerical integration of Eq. (13) with a Lorentzian transmission function (full), constant transmission expression, Eq. (3) with $\tau_0 = 0.61$ (dashed), and anomalous shot noise formula, Eq. (5) (dotted). The vertical dotted line marks the bias $\Delta\mu = 0.5$ eV, corresponding to the bias window in panel (a). Simulations were performed at $T = 10$ K using a symmetric potential drop, $\alpha = 1/2$.

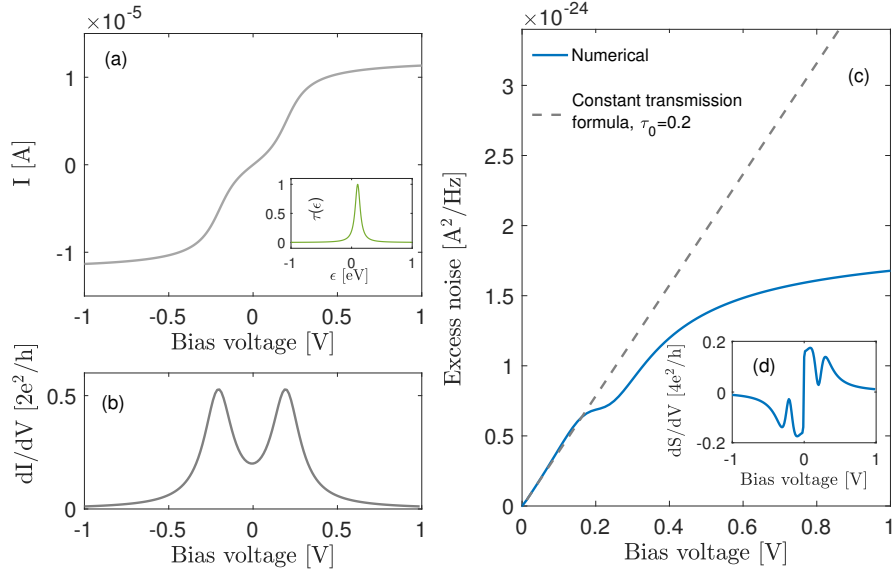

Figure 3: Simulations of current and excess noise using a narrow transmission resonance. (a) Current-voltage characteristics and the Lorentzian transmission function (inset), where $\epsilon_d = 0.1$ eV, $\Gamma = 0.05$ eV. (b) Differential conductance as a function of voltage bias. (c) Excess noise as a function of voltage bias: (full) simulations with Eq. (13) are compared to (dashed) the constant transmission expression of Eq. (3). (d) Differential excess noise illustrating nonlinear trends. Simulations were performed at $T = 10$ K with $\alpha = 1/2$.

In Fig. 3, we depart from our working assumptions and study the case of a narrow resonance, which cannot be captured by Eqs. (2)-(6). Resorting to a direct numerical simulation of Eq. (13), we find that the excess current noise increases linearly at low bias, displays a kink around $\Delta\mu = 0.2$ eV, and approaches saturation around 1 eV. Overall, the noise is concave in voltage and it is highly nonlinear. At the same time, the differential conductance is symmetric in voltage, see panel (b). These concurrent characteristics for the conductance and the noise were observed in some junctions in the experimental work of Ref.,[25] as well as in some of our junction realizations (reported below and in Section S4, Supporting Information). They are reproduced here without the need to invoke many body, polarization effects that show up as e.g. a voltage-dependent resonant level energy, $\epsilon_d(V)$.

## 2.3 Analysis of experimental data

In this Section, we report on measurements of shot noise in atomic-scale Au junctions. At high voltage, the measured noise shows a non-linear behavior with no specific trends and it largely varies between the inspected junctions. Nevertheless, we demonstrate that our formulae for the anomalous shot noise *quantitatively* capture classes of results from this rich data, thus supporting our theoretical picture and modeling.

We use the mechanically-controllable break junction technique at cryogenic conditions (4.2 K) to form atomic-scale junctions and measure the current and its noise. By repeatedly breaking and reforming the junction, an ensemble of different junction realizations with somewhat different atomic configurations are assembled. For each newly-formed junction, differential conductance measurements are performed before and after noise measurements so as to verify the stability of the junction.

Shot noise and conductance measurements at low voltage allow to access the number of conductance channels and their individual contributions.[15] Focusing on the low bias voltage regime, the conductance and noise data of our Au atomic junctions is well explained by decomposing the total transmission into a dominant channel ('1'), along with a minor contribution from a secondary channel ('2'). The latter channel possibly results due to direct tunneling of electrons between the two Au electrodes.

The shot noise data collected in our experiments manifest nonlinear behavior with rich trends. From the ensemble of experimental results, we present four examples in Figs. 4, 5, 6, and 7, showing an enhancement or suppression of shot noise relative to the linear case. Considering Figs. 4-6, these examples reveal distinctive nonlinear trends. However, we selected them based on a common characteristic, that the differential conductance is approximately linear in the bias voltage. In this case, based on Eq. (1), the conducing junction should be modeled with a certain asymmetry, $\alpha \neq 1/2$, and the corresponding current noise is given by Eq. (6). In contrast, the case $d\langle I\rangle/dV \propto V^2$, which is analyzed in Section S4, Supporting Information, is presented in Fig. 7.

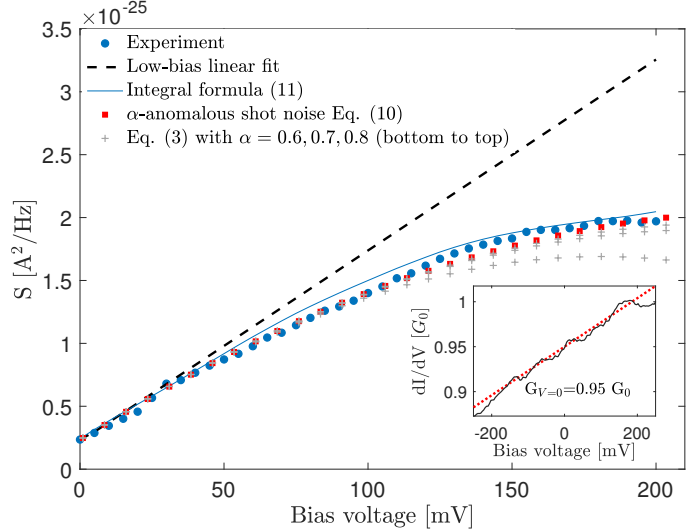

Figure 4: Suppression of shot noise at high voltage. Measurements (○) are compared to a linear fit at low voltage (dashed), and to the $\alpha$-anomalous shot noise formula, Eq. (10) (□) with $T = 6$ K, $\tau_{0,1}$=0.95, $\tau_{0,2}$=0.0064, $A = 0.27$ $G_0/eV$. We further show theoretical curves using Eq. (2) (+) as well as (full) simulations based on Eq. (11). The parameter $A$ is obtained from the differential conductance (inset), with experimental data (full) fitted to a linear line (dotted).

To analyze the data of Figs. 4-6, we assume that the transmission of the dominant channel can be approximated by $\tau_1(\epsilon) \approx \tau_{0,1} + \tau_1'(\mu)(\epsilon - \mu)$. For the secondary channel, it suffices to take into account a constant contribution, $\tau_2(\epsilon) \approx \tau_{0,2}$. The charge current, Eq. (1), follows $\langle I \rangle = \frac{2e}{h}(\tau_{0,1} + \tau_{0,2})\Delta\mu + \frac{2e}{h}\tau_1'(\mu)\left(\alpha - \frac{1}{2}\right)(\Delta\mu)^2$, and the differential conductance satisfies ($\Delta\mu = eV$, $G_0 = \frac{2e^2}{h}$),

$$\frac{d\langle I \rangle}{dV} = G_0(\tau_{0,1} + \tau_{0,2}) + 2G_0\tau_1'(\mu)\left(\alpha - \frac{1}{2}\right)\Delta\mu. \qquad (9)$$

The current noise for setups with $\alpha \neq \frac{1}{2}$ is described by Eq. (6), and we concretize this formula here as (including only the lowest-order nonlinear term),

$$S_{\Delta\mu}^{T\to 0} \approx \sum_{i=1,2} 4k_B T \tau_{0,i} G_0$$
$$+ 4k_B T G_0 \left[\frac{\Delta\mu}{2k_B T} \coth\left(\frac{\Delta\mu}{2k_B T}\right) - 1\right] \sum_{i=1,2} \tau_{0,i}(1-\tau_{0,i})$$
$$+ 2G_0(1-2\tau_{0,1})\tau_1'(\mu)\left(\alpha - \frac{1}{2}\right)(\Delta\mu)^2. \tag{10}$$

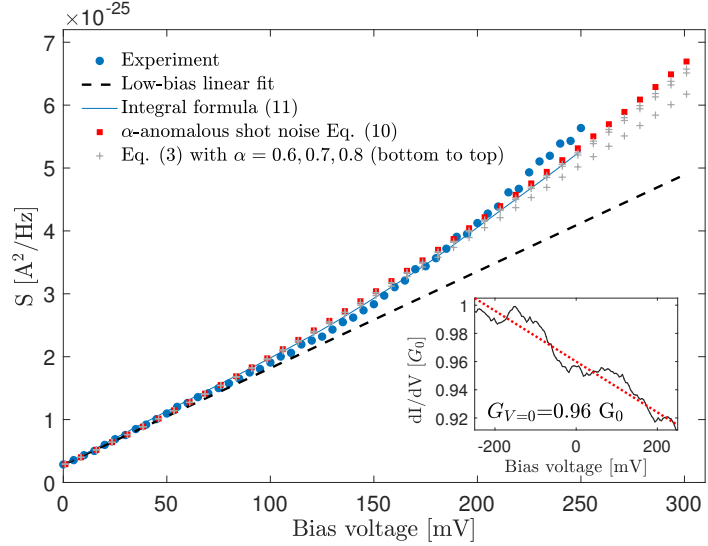

Figure 5: Enhancement of shot noise at high voltage. Measurements (∘) are compared to a linear fit at low voltage (dashed) and to the $\alpha$-anomalous shot noise formula Eq. (10) (□), $T = 7$ K, $\tau_{0,1}$=0.96, $\tau_{0,2}$=0.013, $A = -0.18\ G_0/eV$. We further show theoretical curves using Eq. (2) (+), as well as (full) simulations based on Eq. (11). The parameter $A$ is obtained from the differential conductance (inset), with experimental data (full) fitted to a linear line (dotted).

Equation (10), which we coin the "$\alpha$-anomalous shot noise" formula, is utilized in Figs. 4-6 to recreate experimental data.

Importantly, all parameters building up Eq. (10) can be obtained directly from the experimental differential conductance and the *low-voltage*, linear shot noise data: (i) Focusing on the behavior of the electrical conductance and the shot noise at low voltage we extract the total zero-bias conductance $G = \sum_{i=1,2} G_0 \tau_{0,i}$, and the Fano factor $F = \frac{\sum_{i=1,2} \tau_{0,i}(1-\tau_{0,i})}{\sum_{i=1,2} \tau_{0,i}}$. Under the assumption of two conduction channels, we use these two mathematical relations and isolate the individual contributions, $\tau_{0,1}$ and $\tau_{0,2}$. (ii) The temperature is validated from noise data when approaching $\Delta\mu \to 0$, by fitting the noise to the Johnoson-Nyquist thermal noise expression. (iii) The nonlinear coefficient to the shot noise, $A \equiv 2G_0\tau_1'(\mu)\left(\alpha - \frac{1}{2}\right)$ is gleaned by fitting the differential conductance to a linear function in voltage, see Eq. (9). Altogether, we emphasize that these parameters are gathered from experimental data for the differential conductance and the *linear* shot noise. Next, these parameters are substituted into Eq. (10), and we test whether this simple formula reclaims experimental measurements, which in turn justifies our theoretical modeling.

Figures 4 and 5 exemplify suppression and enhancement, respectively, of the measured shot noise relative to the linear (low bias) limit, which is depicted by a dashed line. We find that the anomalous shot noise formula (10) quantitatively captures these trends in the full range. In both figures, the nonlinear factor is approximately $A \sim \pm 0.2\ G_0/eV$. This value justifies ignoring the next nonlinear term in Eq. (6), which is proportional to $A^2(\Delta\mu)^3$.

So far, we showed that Eq. (10) very well captures the data, thus supporting out modelling. Moreover, we can separately estimate $\tau_1'(\mu)$ and the asymmetry factor $(\alpha - \frac{1}{2})$ by working directly with Eqs. (2)-(4). To achieve that, we vary $\alpha$ in the range 0.55-1 while keeping the product $A \propto \tau_1'(\mu)(\alpha - \frac{1}{2})$ fixed, at the value dictated by the differential conductance. This allows us to estimate the minimal asymmetry in the potential drop that can recover experimental data for the nonlinear shot noise, and we find that $|\alpha - 0.5| \sim 0.2$ typically suffices to explain the measurements. Note that we are unable to differentiate $\alpha = 0.7$ from $\alpha = 0.3$; i.e., we cannot determine the direction of the asymmetry. Curves at larger asymmetry, $\alpha \geq 0.8$, collapse on top of each other.

In Ref.,[25] Tewari and Ruitenbeek suggest that the nonlinearity of shot noise in Au atomic-scale junctions results from quantum interference of electron waves with (randomly placed) defects in the metal contacts. Assuming that the transmission function depends on both energy and voltage, the following formula was developed in Ref.,[25]

$$S(V) = 2eG_0 \sum_i \int_0^V dV' \tau_i(V') \left[1 - \tau_i(V')\right],\qquad(11)$$

with $\sum_i \tau_i(V)$ directly obtained from the differential conductance data, suggested to follow $d\langle I\rangle/dV = G_0 \sum_i \tau_i(V)$. We test this approach on our experimental data by assuming that only the dominant channel depends on voltage, while $\tau_2$ is a constant. We numerically integrate Eq. (11) and find that this expression excellently reproduces our measurements. Note that Eq. (11) was used to describe deviations from a linear dependence of shot noise on voltage in Ref.[25] These deviations were discussed in the context of interference of carriers due to scattering from defects near the junction. However, the derivation of this formula only assumes energy and voltage dependent transmission, which may arise from different mechanisms besides quantum interference. Nevertheless, this numerical approach (i) builds the noise by an explicit integration of the transmission function, (ii) assumes voltage-dependent transmission function, thereby deviating from the principles of the Landauer theory. In contrast, our goal here has been to offer a workable, explicit formula for anomalous shot noise, which is rigorous—derived from the theory of coherent quantum transport. We emphasize that our analytical formulae do not contradict the approach taken in Ref.,[25] but offer further microscopic insights. The variation of the transmission function with energy assumed in our work could be due to quantum interference effects at the contacts as hypothesized in Ref.,[25] and we further expose the impact of voltage drop asymmetry on the transport behavior.

The current noise displayed in Fig. 5 is enhanced relative to the low-voltage linear limit. Can this enhancement result from the activation of a phonon mode at that voltage? Several arguments point against this explanation. First, for pure Au point contacts, previous studies identified phonon density of states up to 20 meV,[43] while we observe deviations from linearity at higher voltage, around 100 mV. The probability for impurities in the junction, such as oxygen (with phonons at 100 meV[44]) is small but finite. However, the enhancement observed here is mild and continuous, while previous explorations of electron-phonon effects in current noise of single atomic-scale junctions showed a more acute kink.[18] Furthermore, the differential conductance that we observe here is an odd function of voltage. In contrast, the activation of vibration should show up as a symmetric step in similar positive and negative bias.[17] All in all, the phonon activation mechanism is inconsistent with our observations.

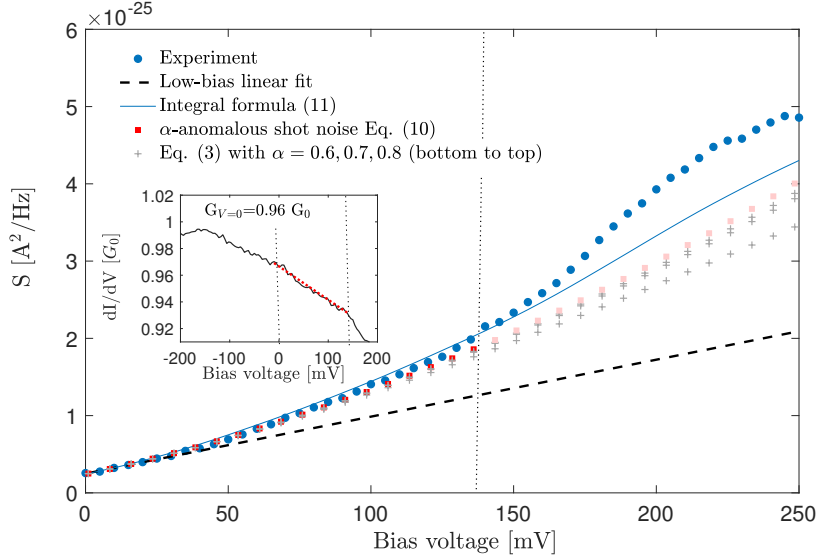

Figure 6: Enhancement of shot noise at high voltage. Shot noise measurements (○) are compared to a linear fit at low voltage (dashed) and to the $\alpha$-anomalous shot noise formula Eq. (10) (□) using $T = 6$ K, $\tau_{0,1}$=0.96, $\tau_{0,2}$=0, $A = -0.25\ G_0/eV$. The parameter $A$ is obtained from the differential conductance (inset), with experimental data (full) fitted to a linear line (dotted). This linear fit is performed between 0 to 140 mV, bounded by the dotted lines. Beyond that, the anomalous shot noise formula (light squares) indeed deviates from experimental data. We further show theoretical curves using Eq. (2) (+) as well as (full line) simulations based on Eq. (11).

We now turn to the third experimental set in Fig. 6. Again, we retrieve from the differential conductance the zero-voltage conductance and the coefficient $A \equiv 2\tau_1'(\mu)\left(\alpha - \frac{1}{2}\right)$. Here, we fitted differential conductance data only below 140 mV receiving $A \sim -0.25\ G_0/eV$; the differential conductance displays more complex trends beyond that, which could stem from the contribution of electron-phonon interaction and $1/f$ noise; the latter is discussed in Section S5, Supporting Information. We use Eq. (10) with its parameters retrieved as explained above, and find that we can faithfully recreate noise measurements up to $\sim 140$

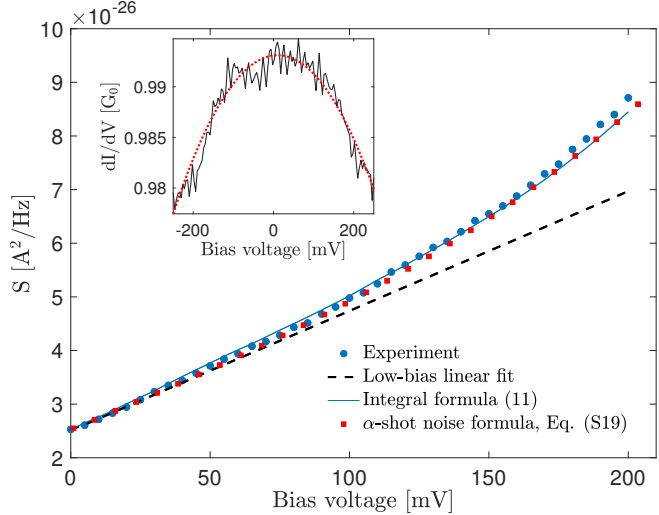

Figure 7: Anomalous shot noise with a quadratic differential conductance-voltage characteristics. Measurements (∘) are compared to a linear fit at low voltage (dashed) and to the anomalous shot noise formula Eq. (S19) (□) using $T = 6$ K and $\tau_{0,1}$=0.992, $\tau_{0,2}$= 0.0010± 0.0011 (we used $\tau_{0,2}$=0.001). We further show (full) simulations based on Eq. (11). Inset: The differential conductance is fitted to a parabola with a curvature $B = -0.231\ G_0/(eV)^2$. This parameter is substituted into Eq. (S19) to recreate the shot noise.

mV.

The analysis presented so far is based on the linear expansion of $\tau(\epsilon)$ with energy around the Fermi energy. In section S4 of the Supporting Information we use a quadratic formula for $\tau(\epsilon)$ and derive an expression for the shot noise [Eq. (S19)], which is parallel to Eqs. (4)-(6). Representative data with a quadratic differential conductance vs. voltage is displayed in Fig. 7. Similarly to the linear expansion, we achieve in the quadratic case a very good theory-experiment agreement. Additional data is presented in Section S4, Supporting Information.

Overall, atomic-scale junctions vary in their transmission probability and bias drop symmetry given differences in the atomic organization in the junction region and the presence of defects in the metal electrodes at the vicinity of the junction. The data presented in this paper was selected after ensuring that the anomalous behavior observed here for shot noise (white noise) is minimally influenced by the low frequency $1/f$ contribution (section S5, Supporting Information). It would be interesting to develop an atomistic model, picture the ensemble of configurations that are generated in atomic-scale junctions, and understand their relation to the generated shot noise at high bias voltage.

# 3 Conclusions

It is often assumed that deviations from the standard shot noise formula, Eq. (3), indicate on the involvement of many-body effects beyond those accounted for in the quantum coherent picture, e.g., electron heating or electron-phonon interaction. However, quantum coherent transport junctions can support 'anomalous' shot noise, an effect that should be assessed before considering additional, complex contributions.

Revisiting the problem of quantum coherent transport, we assumed (i) a weakly energy-dependent transmission function and (ii) a bias drop asymmetry. We then derived a closed-form formula for the shot noise, which is nonlinear at high voltage. Based on the demonstrated theory-experiment agreement, we argue that one should relax the strict assumption of a symmetric voltage bias drop across the junction, even in the seemingly simple Au point contact junction.

Our central results, Eqs. (2) with its limits [Eqs. (5) and (6)] for the anomalous shot noise at low temperature illustrate that: (i) In perfectly symmetric junctions the shot noise develops a cubic contribution at low temperature and high voltage, and the shot noise is always suppressed with respect to low bias voltage. (ii) If the contacts are not identical, the voltage may be divided unevenly, and a quadratic noise term with voltage dominates at high bias. This is manifested by either a suppression or an enhancement of the shot noise relative to the low bias case, depending on the behavior of the transmission function around the Fermi energy.

We derived closed-form formulae for the anomalous shot noise based on the theory of quantum coherent transport. We tested our theory on experimentally obtained data of shot noise in Au atomic-scale junctions and demonstrated a good agreement, which supports the presented modeling. Our analysis further allows us to uncover information on the electronic system: the deviation of the transmission function from a constant value, and the extent of the bias asymmetry across the atomic-scale junction.

What did we achieve? Obviously, if $\tau(\epsilon)$ were entirely known experimentally, the electronic current noise could be obtained directly from a numerical integration of the formal equation (13). However, as a next kin to the transmission function we only have an access to the differential conductance, whose structure at low temperature allows us to deduce on the variation of the transmission function with energy. We showed that based on this information we could explain the behavior of the nonlinear shot noise. This agreement indicates that the origin of the anomalous shot noise is the energy dependent transmission function, which should be taken into account at high bias, as well as bias drop asymmetry, which cannot be ignored at high bias voltage, even in incomplex structures. Our method allows to extract microscopic electronic information on the junction, the transmission constant $\tau(\mu)$, its variation with energy $\tau'(\mu)$, and the bias drop asymmetry—reflecting structural asymmetry. Testing our theory on molecular junctions is therefore an intriguing future direction.

Beyond the quantum coherent limit, strong electron correlations, electron-phonon interactions and other mechanisms for electron scattering should contribute to the appearance of anomalous noise as suggested in Refs.,[25–27] and other studies. The explorations of such effects in single-molecule junctions, to reveal structural and dynamical information is left to future work. Shot noise is the second cumulant of charge fluctuations in steady state; a full-counting statistics analysis recovers all cumulants, and is useful for characterizing the transport process.[46,47] In the context of molecular electronic junctions, understanding the information concealed in high order moments of the current[48] is left to future work. Fundamentally, our theoretical analysis and experimental data can be used to examine the validity of the so-called thermodynamic uncertainty relation, a dissipation-accuracy (fluctuations) tradeoff bound[49] in quantum coherent conductors.[50] Finally, temperature differences across atomic and molecular junctions introduce an additional source of fluctuations as was recently revealed in Ref.[24] The structure of the transmission function further influences this type of noise and it will be analyzed in details in a separate publication.

# 4 Methods

## 4.1 Theory

We consider coherent, elastic transport of electrons in a two-terminal junction. The left ($L$) and right ($R$) metals include collections of noninteracting electrons with occupation numbers following the grand canonical ensemble. The Fermi function $f(\epsilon, \mu_\nu, T) = \frac{1}{e^{\beta(\epsilon-\mu_\nu)}+1}$ is evaluated at the chemical potential $\mu_\nu$ and temperature $T$ with the inverse temperature $\beta = 1/k_B T$, where $k_B$ is the Boltzmann constant; $\nu = L, R$. We denote by $\mu$ the equilibrium Fermi energy. Ignoring decoherence and inelastic processes within the constriction, the average current is given by the Landauer formula,

$$\langle I \rangle = \frac{2e}{h} \int_{-\infty}^{\infty} d\epsilon \, \tau(\epsilon) \left[ f(\epsilon, \mu_L, T) - f(\epsilon, \mu_R, T) \right], \tag{12}$$

with the transmission function $\tau(\epsilon)$. Here, $e$ is the electronic charge. The corresponding zero frequency power spectrum of the noise is given by[2]

$$\begin{aligned}
S &= S_1 + S_2 \\
S_1 &= \frac{4e^2}{h} \int_{-\infty}^{\infty} d\epsilon \{ f(\epsilon, \mu_L, T)[1 - f(\epsilon, \mu_L, T)] + f(\epsilon, \mu_R, T)[1 - f(\epsilon, \mu_R, T)] \} \tau^2(\epsilon), \\
S_2 &= \frac{4e^2}{h} \int_{-\infty}^{\infty} d\epsilon \{ f(\epsilon, \mu_R, T)[1 - f(\epsilon, \mu_L, T)] + f(\epsilon, \mu_L, T)[1 - f(\epsilon, \mu_R, T)] \} \tau(\epsilon)[1 - \tau(\epsilon)].
\end{aligned} \tag{13}$$

In this partition of the total noise, $S_1$ includes additive terms in the left and right metals while $S_2$ collects transport processes from one terminal to the other. The transmission function $\tau(\epsilon)$ is energy dependent; many-body effects (electron-phonon interaction, electronic response to the applied electric field) are sometimes phenomenologically introduced into the transmission

function—though not in our work.

To derive the standard, 'normal' shot noise formula, we approximate $\tau(\epsilon)$ by a constant,[2]

$$
\begin{aligned}
S_{\Delta\mu}^T &= 4k_B T G_0 \sum_i \tau_{0,i}^2 \\
&+ 2\Delta\mu \coth\left(\frac{\Delta\mu}{2k_B T}\right) G_0 \sum_i \tau_{0,i}(1-\tau_{0,i}).
\end{aligned}
\qquad (14)
$$

Here, $\Delta\mu = eV$ is the chemical potential difference due to the bias voltage $V$, $G_0 = 2e^2/h$ is the quantum of conductance. The current and the power noise include contributions from several channels, with $\tau_{0,i}$ the transmission probability of the $i$th channel evaluated at the Fermi energy $\mu$ of the metal electrodes. If the temperature is low relative to the bias, $\coth\left(\frac{|\Delta\mu|}{2k_B T}\right) \to 1$, one recovers the linear relation,

$$
S_{\Delta\mu}^{T\to 0} = 2|\Delta\mu|GF, \qquad (15)
$$

with the Fano factor $F = \sum_i \tau_{0,i}(1-\tau_{0,i})/\sum_i \tau_{0,i}$ and the electrical conductance $G = G_0 \sum_i \tau_{0,i}$.

To explore the impact of an energy dependent transmission function on shot noise, we write down a Taylor expansion for the transmission function, performed around the equilibrium Fermi energy $\mu$,

$$
\tau(\epsilon) \approx \tau(\mu) + \frac{d\tau}{d\epsilon}\bigg|_\mu (\epsilon - \mu). \qquad (16)
$$

This expansion is meaningful as long as the transmission slowly varies with energy within the bias window. For simplicity, we assume a single channel. We identify $\tau(\mu)$ by $\tau_0$ and $\tau'(\mu) \equiv \frac{d\tau}{d\epsilon}\big|_\mu$. Critically, we allow the potential to drop unevenly across the atomic-scale junction,

$$\begin{aligned} \mu_L &= \mu + \alpha \Delta\mu, \\ \mu_R &= \mu - (1-\alpha)\Delta\mu, \end{aligned} \quad (17)$$

with $0 \leq \alpha \leq 1$; when $\alpha = \frac{1}{2}$, the potential drop evenly on the two contacts, $\pm\Delta\mu/2$.

We calculate the current-voltage characteristics of the junction by substituting Eq. (16) into Eq. (12) with the voltage drop Eq. (17). We further derive a closed-form expression for the shot noise by substituting Eq. (16) into Eq. (13), along with the voltage drop Eq. (17). Details are included in Sections S1-S3, Supporting Information. The resulting current is given in Eq. (1) with the current noise (2)-(6).

## 4.2 Experiment

**Formation of Au atomic Junctions.** The mechanically controllable break junction technique[51] in cryogenic temperature is used to form Au atomic junctions. A gold wire (99.99%, 0.1 mm diameter, Goodfellow), with a partial cut in its center is attached to a flexible and insulating substrate. This structure is placed in a vacuum chamber, pumped to $10^{-5}$ mbar and cooled to 4.2 K. The sample is then bent by a piezoelectric element. As a result, the wire is stretched and gradually thinned until a contact with only few atoms down to a single atom in its cross-section is formed between the two wire segments. To measure conductance and noise across the formed atomic junction, the two wire segments are used as electrodes. Repeated squeezing of the electrodes against each other, followed by stretching the reformed contact is used to obtain new atomic junctions. This procedure allows the characterization of an ensemble of atomic junctions with different structures.

**Differential Conductance Measurements.** Differential conductance vs. voltage measurements ($d\langle I\rangle/dV$ vs. $V$) are conducted via a standard lock-in technique, using a Stanford Research SR830 lock-in amplifier. A DC bias voltage signal from a National Instruments (NI) PCI-6221 DAQ card is modulated by an AC voltage produced by the lock-in amplifier (3 mV rms at about 3.33 kHz). The resulting current across the sample is amplified by a current preamplifier (SR570) and sent back to the lock-in to extract the corresponding signal at the frequency of the applied AC modulation. Differential conductance measurements are performed before and after each set of noise measurements in order to verify that the contact maintained its stability during the noise measurement by comparing the two differential conductance spectra.

**Shot Noise Measurements.** Noise measurements are performed on the atomic junctions using a dedicated circuit.[24] To measure noise on atomic junctions, the sample is disconnected from the conductance measurement circuit and connected to the dedicated circuit using switches. The sample is current-biased by a Yokogawa GS200 SC voltage source connected to the sample through two 0.5MΩ resistors located in proximity to the sample. The resulting voltage noise is amplified by a custom-made differential low-noise amplifier and analyzed via a NI PXI-5922 DAQ card, using a LabView implemented fast Fourier transform analysis. For each stable atomic junction, noise measurements are conducted at a set of different bias currents, where at each bias 3,000 measurements of noise spectra are taken and averaged.

## Acknowledgement

DS acknowledges the Natural Sciences and Engineering Research Council (NSERC) of Canada Discovery Grant and the Canada Chairs Program. OT appreciates the support of the Harold Perlman family, and acknowledges funding by a research grant from Dana and Yossie Hollander, the Israel Science Foundation (grant number 1089/15), and the Minerva Foundation (grant number 120865).

# Supporting Information Available

Supporting Information Available: S1. Working formulae; S2. Review of "normal shot noise": Constant transmission function; S3. Derivation of Eqs. (2)-(6): anomalous shot noise; S4. Theory-experiment analysis of other types of junctions; S5. Estimation of the contribution of the $1/f$ noise. This material is available free of charge via the Internet at http://pubs.acs.org.

# Graphical TOC Entry

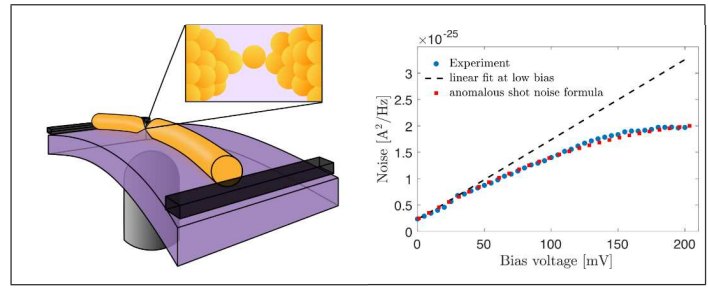

# Supporting Information: Origin of the Anomalous Electronic Shot Noise in Atomic-Scale Junctions

Anqi Mu,[a,†] Ofir Shein Lumbroso[a,‡] Oren Tal,[‡] and Dvira Segal[*,†]

†Department of Chemistry and Centre for Quantum Information and Quantum Control, University of Toronto, 80 Saint George St., Toronto, Ontario, Canada M5S 3H6

‡Department of Chemical and Biological Physics, Weizmann Institute of Science, Rehovot, Israel

E-mail: dvira.segal@utoronto.ca

[a]A.M. and O.S.L. contributed equally to this work

S1. Working formulae

S2. Review of "normal shot noise": Constant transmission function

S3. Derivation of Eq. (2): anomalous shot noise

S4. Theory-experiment analysis of other types of junctions

S5. Estimation of the contribution of $1/f$ noise

# S1 Working formulae

To make our analysis self-contained, in this Section we describe our modeling and theoretical expressions. This information is partially included in the Method section of the main text, but we repeat it here for completeness. We consider coherent, elastic transport of electrons in a two-terminal junction. The metals include collections of noninteracting electrons with occupation numbers following the grand canonical ensemble; the Fermi function $f(\epsilon, \mu_\nu, T) = \frac{1}{e^{\beta(\epsilon-\mu_\nu)}+1}$ is evaluated at the chemical potential $\mu_\nu$ and temperature $T$ with the inverse temperature $\beta = 1/k_B T$; $\nu = L, R$. Below we denote by $\mu$ the equilibrium Fermi energy. Ignoring decoherence and inelastic processes within the constriction, the average current is given by the Landauer formula,

$$\langle I \rangle = \frac{2e}{h} \int_{-\infty}^{\infty} d\epsilon\, \tau(\epsilon) \left[ f(\epsilon, \mu_L, T) - f(\epsilon, \mu_R, T) \right]. \tag{S1}$$

The corresponding zero frequency power spectrum of the noise is given by [1]

$$\begin{aligned}
S &= S_1 + S_2 \\
S_1 &= \frac{4e^2}{h} \int_{-\infty}^{\infty} d\epsilon \{ f(\epsilon, \mu_L, T)[1 - f(\epsilon, \mu_L, T)] + f(\epsilon, \mu_R, T)[1 - f(\epsilon, \mu_R, T)] \} \tau^2(\epsilon), \\
S_2 &= \frac{4e^2}{h} \int_{-\infty}^{\infty} d\epsilon \{ f(\epsilon, \mu_R, T)[1 - f(\epsilon, \mu_L, T)] + f(\epsilon, \mu_L, T)[1 - f(\epsilon, \mu_R, T)] \} \tau(\epsilon)[1 - \tau(\epsilon)].
\end{aligned} \tag{S2}$$

In this partition of the total noise, $S_1$ includes additive terms in the left and right metals while $S_2$ collects transport processes from one terminal to the other. The transmission function $\tau(\epsilon)$ is energy dependent; voltage and temperature dependency, rooted in many-body effects, are sometimes phenomenologically introduced into the transmission function—though not in our work.

To derive the standard result for the shot noise, one assumes a constant transmission function. In this work, we write down a Taylor expansion for the transmission function, performed around the equilibrium Fermi energy $\mu$,

$$\tau(\epsilon) \approx \tau(\mu) + \frac{d\tau}{d\epsilon}\bigg|_\mu (\epsilon - \mu). \tag{S3}$$

For simplicity, we denote $\tau(\mu)$ by $\tau_0$ and $\tau'(\mu) \equiv \frac{d\tau}{d\epsilon}\big|_\mu$. Another central ingredient of our work is that we allow the applied potential to drop asymmetrically around the equilibrium Fermi energy,

$$\begin{aligned}\mu_L &= \mu + \alpha \Delta\mu, \\ \mu_R &= \mu - (1-\alpha)\Delta\mu,\end{aligned} \tag{S4}$$

with $0 \leq \alpha \leq 1$; when $\alpha = 1/2$, the potential bias is partitioned symmetrically at the two ends.

## S2 Review of "normal shot noise": Constant transmission function

Let us now review the standard, 'normal' shot noise expression, which is used to fit experimental observations of shot noise at low voltage. Equations (S1)-(S2) can be simplified if $\tau(\epsilon)$ is assumed a constant. This assumption is justified at low bias voltage. Then, e.g., the width of resonances (responsible for charge transport through the conductor) is considerable relative to the bias window and the transmission function can be approximated by its (fixed) value at the Fermi energy. Making this critical assumption, the averaged current under a finite voltage reduces to $\langle I \rangle = \frac{2e}{h}\Delta\mu \sum_i \tau_{0,i}$, with the power noise[1,2]

$$\begin{aligned}S^T_{\Delta\mu} &= 4k_B T G_0 \sum_i \tau_{0,i}^2 \\ &+ 2\Delta\mu \coth\left(\frac{\Delta\mu}{2k_B T}\right) G_0 \sum_i \tau_{0,i}(1 - \tau_{0,i}).\end{aligned} \tag{S5}$$

S3

Here $\Delta\mu = eV$ is the chemical potential difference due to the bias voltage $V$, $G_0 = 2e^2/h$ is the quantum of conductance. The current and the power noise may include contributions from multiple channels, with $\tau_{0,i}$ the transmission probability of the $i$th channel evaluated at the Fermi energy $\mu$. Eq. (S5) is well known; we retrieve it in Sec. S3 as a special limit of a more general expression.

Low bias measurements of shot noise in atomic-scale and molecular junctions agree well with Eq. (S5), see for example Refs.[3-6] Specifically, when the temperature is low relative to the bias, $\coth(\frac{|\Delta\mu|}{2k_BT}) \to 1$, and we get

$$S_{\Delta\mu}^{T\to 0} = 2|\Delta\mu|GF. \tag{S6}$$

Here, $F = \sum_i \tau_i(1-\tau_{0,i})/\sum_i \tau_{0,i}$ is the Fano factor, $G = G_0 \sum_i \tau_{0,i}$ stands for the electrical conductance. The noise (S6) is linear in voltage. Therefore, nonlinearity of the shot noise at high voltage corresponds to an 'anomalous' behavior. Since $\langle I \rangle = GV$ and $\Delta\mu = eV$, we can organize Eq. (S6) in its familiar form as $S_{\Delta\mu}^{T\to 0} = 2e|\langle I \rangle|F$.

We now consider a junction at equilibrium, $\Delta\mu = 0$. Eq. (S2) then reduces to the Johnson-Nyquist thermal noise,

$$S_{\Delta\mu=0}^{T} = 4k_BTG, \tag{S7}$$

with the electrical conductance $G = \frac{2e^2}{h}\int d\epsilon\, \tau(\epsilon)\left(-\frac{df}{d\epsilon}\right)$. Note that we can also approach the equilibrium limit from Eq. (S5) and arrive at a corresponding result. Nevertheless, Eq. (S7) holds without assuming a constant transmission function.

## S3 Derivation of Eqs. (2)-(6): anomalous shot noise

We examine here the behavior of shot noise under high voltage; we assume that there is no applied temperature difference. We begin by evaluating $S_1$ in Eq. (S2) using the linear expansion for the transmission function, Eq. (S3). For convenience, we assume a single channel. We omit the prefactor $\frac{4e^2}{h}$ and re-install it only at the end of our derivation,

$$S_1 = \int_{-\infty}^{\infty} d\epsilon \left( -k_B T \frac{\partial f_L}{\partial \epsilon} - k_B T \frac{\partial f_R}{\partial \epsilon} \right) \times [\tau_0 + \tau'(\mu)(\epsilon - \mu)]^2. \tag{S8}$$

Here, $f_\nu = f(\epsilon, \mu_\nu, T)$, $\Delta \mu = \mu_L - \mu_R$, $\mu_L = \mu + \alpha \Delta \mu$, $\mu_R = \mu - (1-\alpha)\Delta \mu$ and $T = T_L = T_R$. Explicitly,

$$\begin{aligned} S_1 &= \int_{-\infty}^{\infty} d\epsilon \left\{ (-k_B T) \left[ \tau_0^2 \frac{\partial f_L}{\partial \epsilon} + 2\tau_0 \tau'(\mu)(\epsilon - \mu) \frac{\partial f_L}{\partial \epsilon} + [\tau'(\mu)]^2 (\epsilon - \mu)^2 \frac{\partial f_L}{\partial \epsilon} \right] \right. \\ &+ \left. (-k_B T) \left[ \tau_0^2 \frac{\partial f_R}{\partial \epsilon} + 2\tau_0 \tau'(\mu)(\epsilon - \mu) \frac{\partial f_R}{\partial \epsilon} + [\tau'(\mu)]^2 (\epsilon - \mu)^2 \frac{\partial f_R}{\partial \epsilon} \right] \right\}. \end{aligned} \tag{S9}$$

We now evaluate the different terms,

$$\begin{aligned} I_1 &\equiv \int_{-\infty}^{\infty} d\epsilon (-k_B T) \tau_0^2 \frac{\partial f_L}{\partial \epsilon} = k_B T \tau_0^2, \\ I_2 &\equiv \int_{-\infty}^{\infty} d\epsilon (-k_B T) 2\tau_0 \tau'(\mu) \left[ \epsilon - (\mu_L - \alpha \Delta \mu) \right] \frac{\partial f_L}{\partial \epsilon} \\ &= \int_{-\infty}^{\infty} d\epsilon (-k_B T) 2\tau_0 \tau'(\mu) (\epsilon - \mu_L) \frac{\partial f_L}{\partial \epsilon} + \int_{-\infty}^{\infty} d\epsilon (-k_B T) 2\tau_0 \tau'(\mu) \alpha \Delta \mu \frac{\partial f_L}{\partial \epsilon} \\ &= 2k_B T \tau_0 \tau'(\mu) \alpha \Delta \mu, \\ I_3 &\equiv \int_{-\infty}^{\infty} d\epsilon (-k_B T) [\tau'(\mu)]^2 \left[ \epsilon - (\mu_L - \alpha \Delta \mu) \right]^2 \frac{\partial f_L}{\partial \epsilon} \\ &= \int_{-\infty}^{\infty} d\epsilon (-k_B T) [\tau'(\mu)]^2 \left[ (\epsilon - \mu_L)^2 + 2\alpha \Delta \mu (\epsilon - \mu_L) + \alpha^2 (\Delta \mu)^2 \right] \frac{\partial f_L}{\partial \epsilon} \\ &= k_B T [\tau'(\mu)]^2 \frac{\pi^2 k_B^2 T^2}{3} + k_B T [\tau'(\mu)]^2 \alpha^2 (\Delta \mu)^2. \end{aligned} \tag{S10}$$

Summing up these integrals, along with the corresponding contributions from the right side, we get

$$\begin{aligned} S_1 &= 2k_B T \tau_0^2 + 2k_B T \tau_0 \tau'(\mu) \Delta \mu \alpha - 2k_B T \tau_0 \tau'(\mu) \Delta \mu (1 - \alpha) \\ &+ 2k_B T [\tau'(\mu)]^2 \frac{\pi^2 k_B^2 T^2}{3} + k_B T [\tau'(\mu)]^2 \left[ \alpha^2 (\Delta \mu)^2 + (1-\alpha)^2 (\Delta \mu)^2 \right]. \end{aligned} \tag{S11}$$

Next, we evaluate $S_2$ in Eq. (S2). Under bias voltage it can be organized as

$$S_2 = \coth\left(\frac{\Delta\mu}{2k_BT}\right) \int_{-\infty}^{\infty} d\epsilon \, [f_L(\epsilon) - f_R(\epsilon)] [\tau_0 + \tau'(\mu)(\epsilon - \mu)][1 - \tau_0 - \tau'(\mu)(\epsilon - \mu)]. \tag{S12}$$

The integral can be evaluated *exactly* using the following relations,

$$\begin{aligned}
I_4 &\equiv \coth\left(\frac{\Delta\mu}{2k_BT}\right) \int_{-\infty}^{\infty} d\epsilon [f_L(\epsilon) - f_R(\epsilon)] \tau_0(1-\tau_0) = \tau_0(1-\tau_0)\Delta\mu \coth\left(\frac{\Delta\mu}{2k_BT}\right), \\
I_5 &\equiv \coth\left(\frac{\Delta\mu}{2k_BT}\right) \int_{-\infty}^{\infty} d\epsilon [f_L(\epsilon) - f_R(\epsilon)](1-2\tau_0)\tau'(\mu)(\epsilon - \mu) \\
&= \coth\left(\frac{\Delta\mu}{2k_BT}\right)(1-2\tau_0)\tau'(\mu) \int_{-\infty}^{\infty} d\epsilon [f_L(\epsilon) - f_R(\epsilon)] \left\{\left[\epsilon - \left(\mu + \left(\alpha - \frac{1}{2}\right)\Delta\mu\right)\right] + \left(\alpha - \frac{1}{2}\right)\Delta\mu\right\} \\
&= \coth\left(\frac{\Delta\mu}{2k_BT}\right)(1-2\tau_0)\tau'(\mu) \left[\alpha - \frac{1}{2}\right](\Delta\mu)^2, \\
I_6 &\equiv \coth\left(\frac{\Delta\mu}{2k_BT}\right) \int_{-\infty}^{\infty} d\epsilon [f_L(\epsilon) - f_R(\epsilon)][\tau'(\mu)]^2 \left\{\left[\epsilon - \left(\mu + \left(\alpha - \frac{1}{2}\right)\Delta\mu\right)\right] + \left(\alpha - \frac{1}{2}\right)\Delta\mu\right\}^2 \\
&= \coth\left(\frac{\Delta\mu}{2k_BT}\right)[\tau'(\mu)]^2 \left[\Delta\mu \frac{\pi^2 k_B^2 T^2}{3} + \frac{1}{12}(\Delta\mu)^3 + \left(\alpha - \frac{1}{2}\right)^2 (\Delta\mu)^3\right]. 
\end{aligned} \tag{S13}$$

Overall, we get

$$\begin{aligned}
S_2 &= \tau_0(1-\tau_0)\Delta\mu \coth\left(\frac{\Delta\mu}{2k_BT}\right) + \coth\left(\frac{\Delta\mu}{2k_BT}\right)(1-2\tau_0)\tau'(\mu)\left(\alpha - \frac{1}{2}\right)(\Delta\mu)^2 \\
&\quad - \coth\left(\frac{\Delta\mu}{2k_BT}\right)[\tau'(\mu)]^2 \left[\Delta\mu \frac{\pi^2 k_B^2 T^2}{3} + \frac{1}{12}(\Delta\mu)^3 + \left(\alpha - \frac{1}{2}\right)^2 (\Delta\mu)^3\right]. 
\end{aligned} \tag{S14}$$

Combining $S_1$ [Eq. (S11)] and $S_2$, we get the voltage-activated anomalous shot noise,

$$\begin{aligned} S_{\Delta\mu}^T &= 2k_B T \tau_0^2 + 2k_B T \tau_0 \tau'(\mu) \Delta\mu \alpha - 2k_B T \tau_0 \tau'(\mu) \Delta\mu (1-\alpha) \\ &+ 2k_B T [\tau'(\mu)]^2 \frac{\pi^2 k_B^2 T^2}{3} + k_B T [\tau'(\mu)]^2 \left[\alpha^2 (\Delta\mu)^2 + (1-\alpha)^2 (\Delta\mu)^2\right] \\ &+ \tau_0 (1-\tau_0) \Delta\mu \coth\left(\frac{\Delta\mu}{2k_B T}\right) + \coth\left(\frac{\Delta\mu}{2k_B T}\right)(1-2\tau_0)\tau'(\mu)\left(\alpha - \frac{1}{2}\right)(\Delta\mu)^2 \\ &- \coth\left(\frac{\Delta\mu}{2k_B T}\right)[\tau'(\mu)]^2 \left[\Delta\mu \frac{\pi^2 k_B^2 T^2}{3} + \frac{1}{12}(\Delta\mu)^3 + \left(\alpha - \frac{1}{2}\right)^2 (\Delta\mu)^3\right]. \end{aligned} \quad (S15)$$

Multiplying it by $2G_0 = \dfrac{4e^2}{h}$ we obtain Eqs. (2)-(4) in the main text. It is significant to note that this result is exact in $\Delta\mu$, to the order of $\tau'(\mu)$ considered.

## S4 Theory-experiment analysis of other types of junctions

So far, our focus has been on junctions whose transmission function can be approximated by the expansion (S3) around the Fermi energy. This expansion leads to a differential conductance that is linear in voltage. In this section, we consider an alternative form for the transmission function, relevant for a complementary class of junctions,

$$\tau(\epsilon) \approx \tau_0 + \frac{1}{2}\tau''(\mu)(\epsilon - \mu)^2. \quad (S16)$$

We again allow the voltage bias to drop asymmetrically around the equilibrium Fermi energy, $\mu_L = \mu + \alpha\Delta\mu$, $\mu_R = \mu - (1-\alpha)\Delta\mu$, with $0 \leq \alpha \leq 1$. Using the Landauer formula we obtain the averaged charge current as,

$$\langle I \rangle = \frac{2e}{h}\tau_0 \Delta\mu + \frac{e}{h}\tau''(\mu)\left[\Delta\mu \frac{\pi^2 k_B^2 T^2}{3} + \frac{1}{12}(\Delta\mu)^3 + \left(\alpha - \frac{1}{2}\right)^2 (\Delta\mu)^3\right]. \quad (S17)$$

We now assume that the temperature is low relative to the bias voltage and derive the differential conductance,

$$\frac{d\langle I\rangle}{dV}\bigg|_{T\to 0} = G_0\tau_0 + \frac{3G_0}{2}\tau''(\mu)\left[\frac{1}{12} + \left(\alpha - \frac{1}{2}\right)^2\right](\Delta\mu)^2. \tag{S18}$$

By fitting the differential conductance to a parabola we obtain the curvature $B \equiv \frac{3}{2}G_0\tau''(\mu)[1/12 + (\alpha-1/2)^2]$, with $\frac{d\langle I\rangle}{dV}|_{T\to 0} = G_0\tau_0 + B(\Delta\mu)^2$. Note that the quadratic term in $\frac{d\langle I\rangle}{dV}$ is sustained even when $\alpha = 1/2$.

Repeating the procedure of Sec. S3, we derive a closed form formula for the current noise using the transmission function (S16). Here, we present only the low temperature limit,

$$\begin{aligned}S_{\Delta\mu}^{T\to 0} &= 4G_0 k_B T \tau_0^2 + 2G_0\tau_0(1-\tau_0)\Delta\mu\coth\left(\frac{\Delta\mu}{2k_B T}\right) \\ &+ 2G_0(1-2\tau_0)\frac{\tau''(\mu)}{2}\left[\frac{1}{12} + \left(\alpha - \frac{1}{2}\right)^2\right]|(\Delta\mu)|^3\end{aligned} \tag{S19}$$

It is significant to note that the curvature $B$, which determines the nonlinearity of the differential conductance, dictates the anomalous component of the shot noise. Nevertheless, in the present model, Eq. (S16), we cannot in practice determine the extent of bias asymmetry $\alpha$.

We illustrate the analysis and the validity of Eq. (S19) on experimental data of shot noise in Au atomic junctions. Besides results in the main text, Figure S1 displays data for which the differential conductance is approximately quadratic around zero voltage (inset), which make it suitable for the present analysis. Following our procedure, we extract the curvature of the parabola $B$ from the differential conductance and employ it in Eq. (S19) to generate the current noise. In Fig. 7 (main text), the quadratic behavior of the differential conductance with voltage extends up to 200 mV, and the noise is indeed well reproduced throughout the whole range. In contrast, in Fig. S1, the quadratic behavior only extends up to 120 mV, beyond which deviations show; in accordance, we properly capture the experimental data for the shot noise up to this voltage. Remarkably, in Fig. S1 the nonlinear (cubic) term almost immediately dominates at low voltage, since the curvature is quite large. Finally, Fig. S2 displays differential conductance data that does not show a definite quadratic trend (extending the experiment to higher voltage could strengthen this model). Nevertheless, we test the quadratic formula on this data and show that we qualitatively capture the overall trend of the experimental shot noise, observing an enhancement of noise relative to the low bias case.

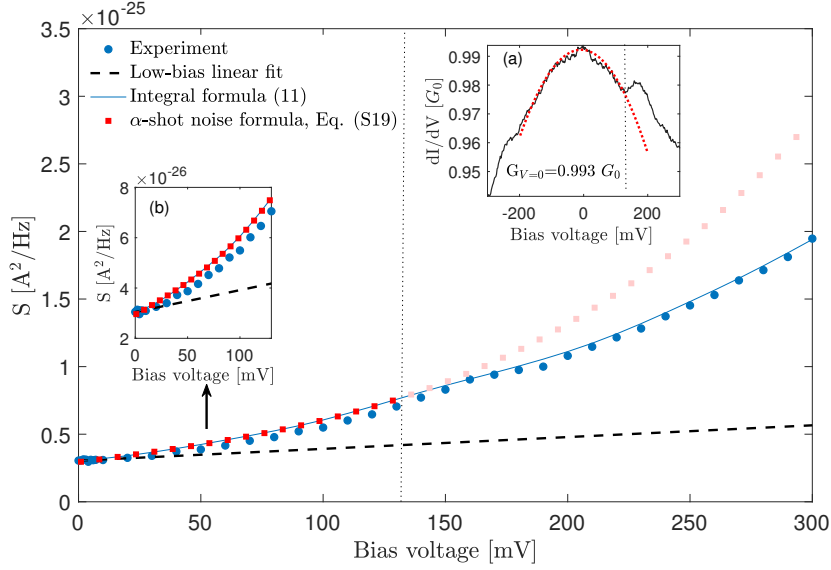

**Figure S1:** Anomalous shot noise with a quadratic differential conductance-voltage characteristics. Measurements (○) are compared to a linear fit at low voltage (dashed) and to the anomalous shot noise formula Eq. (S19) (□) using $T = 7$ K, $\tau_{0,1}=0.99$, $\tau_{0,2}=0.00$. We further show (full) simulations based on Eq. (11). (a) We fit the differential conductance data (full) to a parabola (dotted) within the region -200 to 120 mV and get the curvature $B = -0.826$ $G_0/(eV)^2$, which is substituted into Eq. (S19) to reproduce the nonlinear shot noise (b), showing an excellent agreement. Results from the nomalous shot noise formula outside the proper fitting window are further displayed in the main plot (light □).

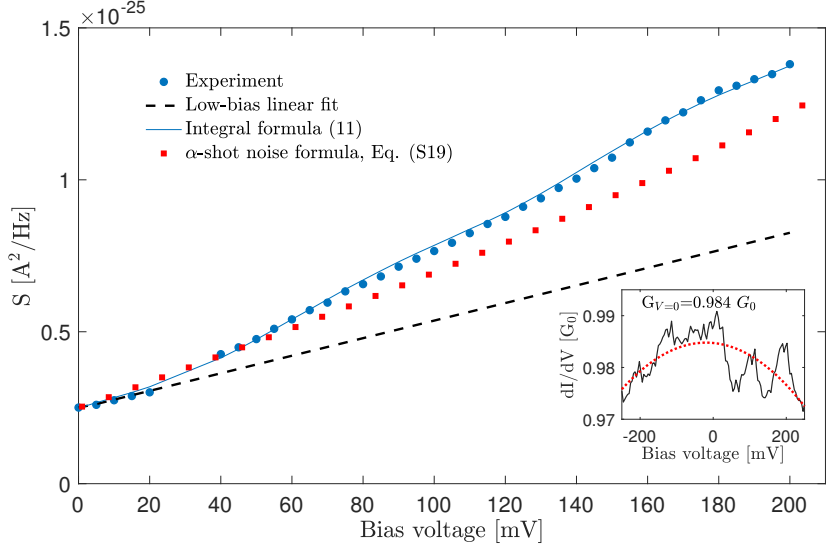

**Figure S2:** Anomalous shot noise with a possible quadratic differential conductance-voltage characteristics. Measurements (○) are compared to the anomalous shot noise formula Eq. (S19) (□) using $T = 6$ K, $\tau_{0,1}=0.984$, $\tau_{0,2}=0.0018$. We further show (full) simulations based on Eq. (11). Inset: We fit the differential conductance (full) to a parabola (dotted) and get the curvature $B = -0.169$ $G_0/(eV)^2$, which is substituted into Eq. (S19) to reproduce the nonlinear shot noise.

## S5  Estimation of the contribution of the $1/f$ noise

At high voltage, other noise sources could contribute to the anomalous behavior. Specifically, the $1/f$ noise grows quadratically with voltage, and its contribution could become significant. As representative examples, in Figs S3 and S4 we display the power spectra of the noise corresponding to Figs. 4 and 6 in the main text. At low frequency, the power spectra shows the $1/f$ noise. The white noise component, which comprises the thermal noise and shot noise is taken in the region $f \approx 1 \times 10^5$ Hz.

We assess the contribution of the $1/f$ component in an approximate manner as follows. First, at each applied voltage we subtract the mean white noise value, to identify the 'pure' $1/f$ contribution. Next, plotting this data on a log-log scale, we extract the power $\alpha$, $S(f) = S_c/f^\alpha$, with $S_c$ a prefactor, which depends on the applied voltage.

Performing this analysis on the data presented in Fig. S3, we obtain the exponent, which somewhat varies with voltage, $\alpha = 1.5 - 1.7$. Since we are interested in the contribution of $1/f$ noise at high voltage, we use $\alpha = 1.66$ at 200 mV. To extract $S_c$ at 200 mV we focus on the low frequency region, and use e.g. the measured value of $S(f = 5000 \text{ Hz}) = 2.3 \times 10^{-17}$ V$^2$/Hz and the calculated power $\alpha = 1.66$. This results in $S_c \sim 3.2 \times 10^{-11}$ V$^2$/Hz. We can now estimate the contribution of the $1/f$ noise at higher frequencies, in what we identify as white noise, say at $f = 9 \times 10^4$ Hz,

$$\begin{aligned} S(f) &= S_c/f^\alpha \\ &= 3.2 \times 10^{-11} \times G^2 \times (9 \times 10^4)^{-1.66} = 1.03 \times 10^{-27} \text{A}^2/\text{Hz}, \end{aligned} \quad (S20)$$

where we used $G = 0.95\, G_0$. The anomalous noise presented in Fig. 4 reaches $2 \times 10^{-25}$ A$^2$/Hz at high voltage, and we conclude that the $1/f$ noise contributes less than 1% to this value.

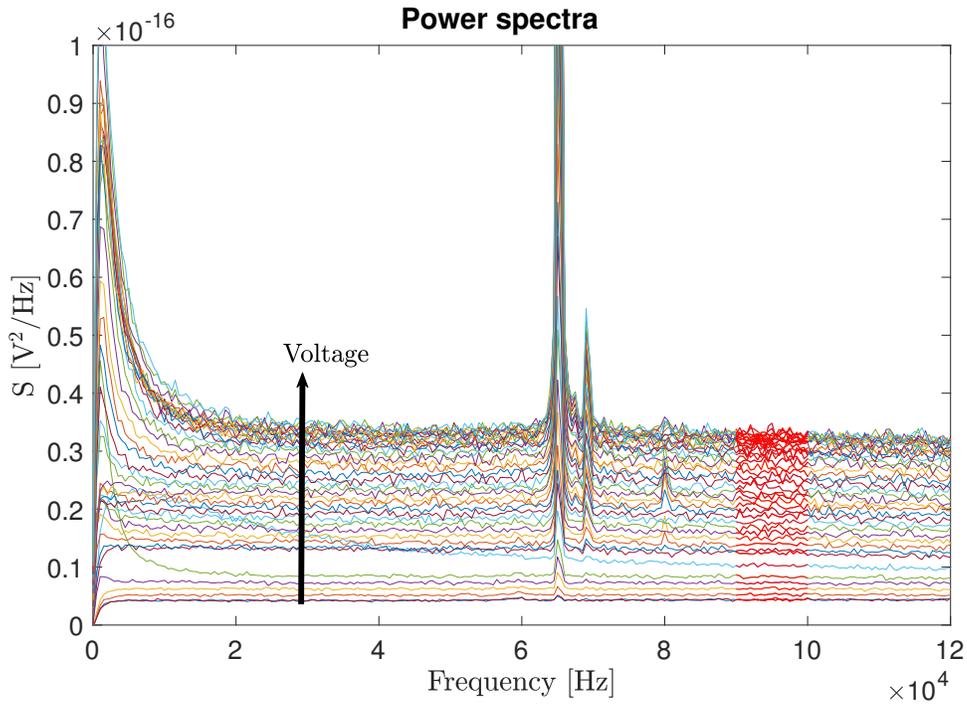

**Figure S3:** Power spectra for different applied voltage. The red region marks the portion of the noise used in the shot noise analysis, leading to Fig. 4 in the main text.

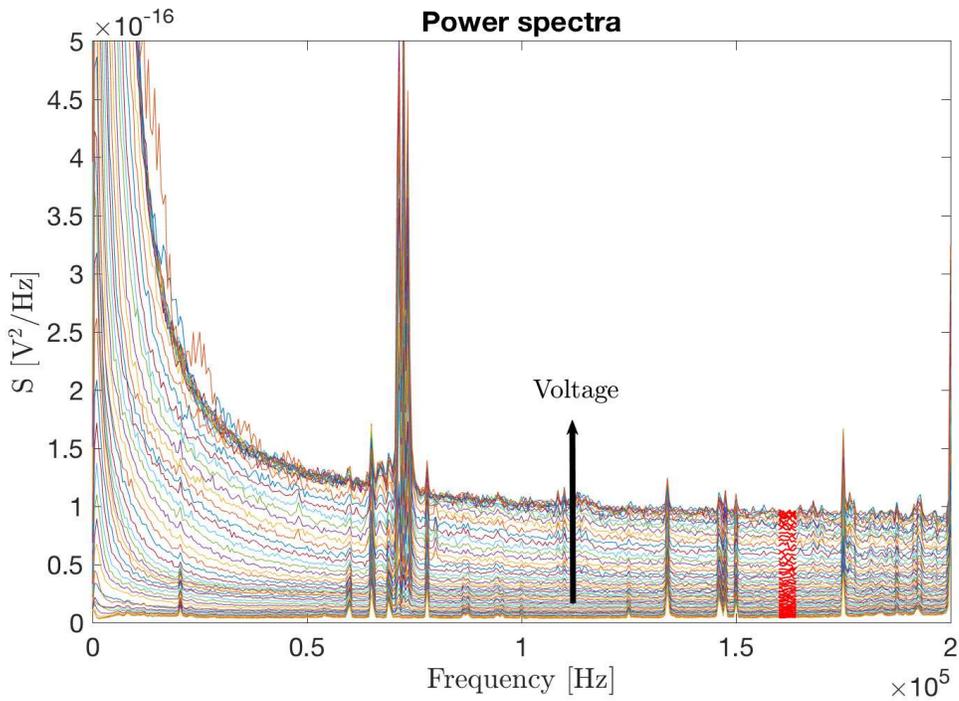

**Figure S4:** Power spectra for different applied voltage. The red region marks the portion of the noise used in the shot noise analysis, leading to Fig. 6 in the main text.

A similar analysis is performed on the power spectra in Fig. S4, which corresponds to Fig. 6 in the main text. Here, at 240 mV we obtain the power $\alpha \sim 1.7$ and the coefficient $S_c = 2.66 \times 10^{-9}$ V$^2$/Hz, which leads to the residual $1/f$ noise $S(f = 1.6 \times 10^5$ Hz$) \sim 2.08 \times 10^{-26}$ A$^2$/Hz, using $G=0.96\, G_0$. In contrast to the former example where the $1/f$ noise is negligible, here we estimate that the $1/f$ noise contributes $\sim$10% to the noise at high voltage, beyond 200 mV. However, in all other cases we confirmed that the $1/f$ noise level was minor, less than 3%, and that the anomalous noise examined was indeed white with negligible frequency-dependent contributions.